\documentclass[twocolumn,showpacs,preprintnumbers,amsmath,amssymb]{revtex4-2}

\usepackage[colorlinks=true,citecolor=blue,urlcolor=blue,linkcolor=blue,bookmarksopen]{hyperref}
\usepackage{graphicx}
\usepackage{lineno}
\usepackage{siunitx}
\usepackage{amsmath}
\usepackage{gensymb}

\begin{document}

\title{Coexistence of Surface Lattice Resonances and Bound states In the Continuum in a Plasmonic Lattice}

\author{Quoc Trung Trinh$^1$}
\author{Sy Khiem Nguyen$^1$}
\author{Dinh Hai Nguyen$^1$}
\author{Gia Khanh Tran$^1$}
\author{Viet Hoang Le$^1$}
\author{Hai-Son Nguyen$^{2,3}$}
\author{Quynh Le-Van$^1$}
\email{quynh.lv@vinuni.edu.vn}

\affiliation{$^1$College of Engineering and Computer Science, VinUniversity, Gia Lam district, Hanoi 14000, Vietnam}
\affiliation{$^2$Univ Lyon, Ecole Centrale de Lyon, CNRS, INSA Lyon, Université Claude Bernard Lyon 1, CPE Lyon, CNRS, INL, UMR5270,
69130 Ecully, France}
\affiliation{$^3$Institut Universitaire de France  (IUF), Paris, France}

\date{\today}

\begin{abstract}
We present a numerical study on a 2D array of plasmonic structures covered by a subwavelength film. We explain the origin of surface lattice resonances (SLRs) using coupled dipole approximation and show that the diffraction-assisted plasmonic resonances and formation of bound states in the continuum (BICs) can be controlled by altering the optical environment. Our study shows that when the refractive index contrast $\Delta$n < -0.1, the SLR cannot be excited, while a significant contrast ($\Delta$n > 0.3) not only sustains plasmonic-induced resonances but also forms both symmetry-protected and accidental BICs. The results can aid the streamlined design of plasmonic lattices in studies on light–matter interactions and applications in biosensors and optoelectronic devices.
\end{abstract}

\maketitle



Surface lattice resonances (SLRs) are delocalized modes emerging as localized surface plasmon resonances (LSPRs) of individual plasmonic particles that are radiatively coupled to the diffraction orders of the incident light waves induced by the lattices  \cite{Zou2004,Markel2005}. The lattice mode is also studied in lossless systems \cite{Babicheva2017}. The coupling is a mechanism to mitigate the ohmic losses from the LSPRs to the diffracted waves. As a result, SLRs are characterized by an increased quality factor over LSPRs and by a sharper Fano resonance in the scattering spectra \cite{Miroshnichenko2010,LeVan2019,BinAlam2021}. These remarkable properties have attracted significant interest for potential applications in lasers \cite{Zhou2013,Yang2015,Ramezani2018}, strong light-matter interactions \cite{Hakala2018,wang2019}, nonlinear optics \cite{Czaplicki2016}, and biosensors \cite{Li2021}. Another peculiar aspect of Fano resonances is their disappearance in the scattering spectra. The collapse of Fano resonances is a signature of the bound state in the continuum (BIC) \cite{Neumann1929,Friedrich1985}. This disappearance leads to an infinite Q-factor and has been experimentally proven in dielectric systems \cite{Hsu2013,Kodigala2017,Ha2018}. Recently, several groups have theoretically proposed \cite{Marinica2008,Azzam2018,MermetLyaudoz2019,Contractor2020,Hsu2021,Wang2020} and reported the BICs in lossy media \cite{Doeleman2018,Liang2020,Seo}. Interestingly, both SLRs and BICs can be realized in periodic structures despite having opposite features in terms of the Fano resonances. To date, there has been no report on any plasmonic system that supports both SLRs and BICs. Here, we propose a simple plasmonic structure made of a silver nanodisks array that can host both SLRs and BICs simultaneously. We elaborate on the physics of the system through numerical studies under varied conditions of the excitation angle, periodicity, and refractive index of the covered layer. We show that three distinct regions with different physics can be obtained for the given system.

\begin{figure}[t]
\centering
\includegraphics[width=0.93\linewidth]{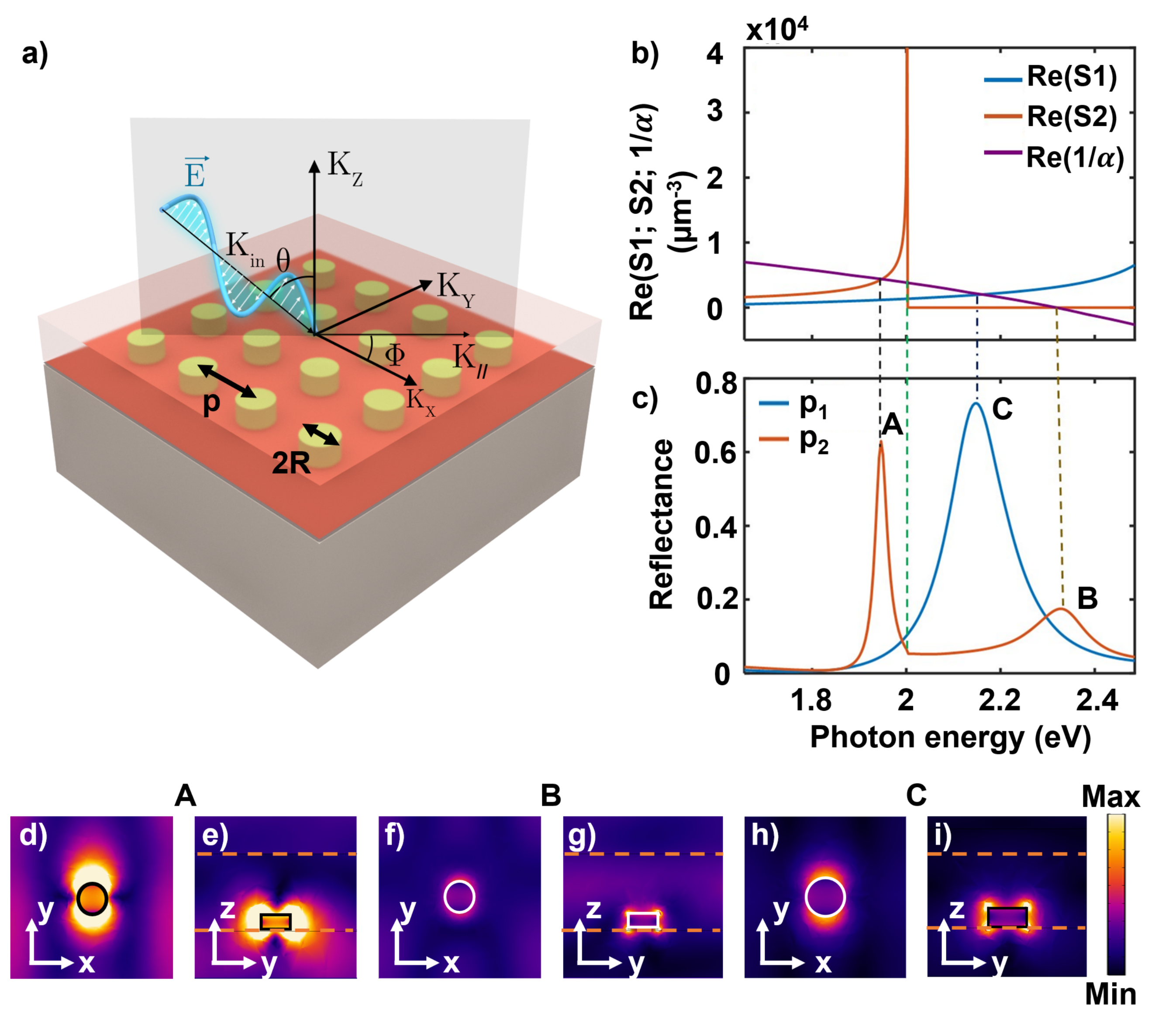}
\caption{The origin of SLR. a) Sketch of the plasmonic array. A square lattice of silver nanodisks on a substrate ($n_{sub}$ = 1.46) is covered by a thin layer. b) The calculated inverse polarizability (1/$\alpha$) of the nanodisk ($r = h = $ 40\,nm) and lattice sums $S_1$ of $p_{1} =$ 325\,nm and $S_2$ of $p_{2} =$ 425\,nm at $\theta = 0^{\circ}$. c) Simulated reflectances at $\theta = 0^{\circ}$ of the structures with $n_{sup}$ = 1.5. Three peaks in $\bold{c)}$ agree well with the crossing points in $\bold{b)}$. The electric fields of peaks: A (d-e), B (f-g) and C (h-i) for a unit cell. The orange dash lines indicate the interfaces between layers.}
\label{fig:fig1}
\end{figure}

We begin our study by addressing the origin of the SLRs in a 2D square lattice at normal incidence. Our system comprises an array of silver nanodisks  (radius $r$ = 40\,nm and height $h$ = 40\,nm) with periodicity $p$, as illustrated in Fig. \ref{fig:fig1}a. They are placed on a glass substrate ($n_{sub}$ = 1.46) and covered by a 200\,nm thick superstrate with refractive index $n_{sup}$. We used the coupled dipole approximation (CDA) to calculate the spectral positions of LSPRs, Rayleigh anomalies, and SLRs \cite{GarciadeAbajo2007}. In this framework, when a periodic array of identical subwavelength nanoparticles is impinged by an external plane wave, each particle can be characterized as an electric dipole with polarizability $\alpha_{E}$\footnote{ $1/\alpha_{E} = \frac{3}{V}\left(L + \frac{\epsilon_{m}}{\epsilon - \epsilon_{m}} \right)$ where V is the volume of the particle, L is the geometric factor that varies with different shapes \cite{Barnes2016}, $\epsilon_{m}$ is the relative permittivity of the medium and $\epsilon(\omega) = 1 - \frac{\omega_{p}^2}{\omega(\omega + i \gamma)}$ with $\omega_p$ being the plasmon frequency  and $\gamma$ is the linewidth of the LSPR.} 
and subjected to the total applied fields, including the incident waves and retarded fields by the neighboring particles. Thus the effective polarizability $\alpha$ of each particle is given by $\alpha^{-1}=\alpha_{E}^{-1} - S$ \footnote{$S \approx \left(\frac{4\sqrt{2}\pi^2}{\sqrt{\beta}} - 118\right)\frac{1}{p^3} + 2i\left(\frac{\pi k}{p^2}-\frac{k^3}{3}+\frac{\pi^2}{(\pi N)^{3/2}\beta^2p^3}\right)$ where $k=(2\pi/\lambda)\sqrt{\epsilon_{r}}$ is the amplitude of wavevector in the medium with relative permittivity $\epsilon_{r}$, $N$ is the number of the particle in the array, and $\beta=\frac{2\pi}{kp} - 1$\cite{Spackova2013}}. Here $S$ is the lattice sum describing the dipole coupling mechanisms between a particle and its neighbors. Note that $\alpha_{E}$ and $S$ are complex values due to the presence of Ohmic and radiative losses. In addition to dipole interactions, the diffraction of the periodic array is another pronounced effect in response to the incident waves. The diffracted waves can radiate in the regions above, possibly below, and in the array. An interesting physical phenomena occur as the wave diffracts in the plane of the array, known as Rayleigh anomaly. The corresponding condition for such an effect is the zero-propagation constant in the propagation direction. In other words, the condition for Rayleigh anomalies in the Cartesian system is given by $\epsilon_r = \left(\sin\theta \cos\Phi + m \frac{\lambda}{p}\right)^2 + \left(\sin\theta \sin\Phi + n \frac{\lambda}{p} \right)^2$ where  the pair $(m,n)$ corresponds to the diffraction orders, and $\theta$ and $\Phi$ are the incident and azimuthal angle, respectively (see Fig. \ref{fig:fig1}a). The far-field reflectance for normal incidence can be expressed as $R = |\frac{2\pi k}{p^2}\alpha|^2$\cite{GarciadeAbajo2007,Teperik2012}, leading to photonic resonances at the zeros of the real part of $\alpha^{-1}$. Figure \ref{fig:fig1}b shows the calculated $\Re(\alpha_E^{-1})$ and real values of the lattice sum $S_1$ and $S_2$ corresponding to two periods $p_{1}$ = 325\,nm and $p_{2}$ = 425\,nm. The $\Re(S_1)$ value is low, indicating a modest contribution of the lattice in the spectral range. In contrast, the  $\Re(S_2)$ shows an abrupt drop to zero-value at 2.0 eV. The intersection of $\Re(\alpha_{E}^{-1})$ with $\Re(S_1)$ is evidence of an LSPR, whereas it crosses $\Re(S_2)$ at three distinct values. 

First, the intersection with the sharp drop (2.0 eV) corresponds to the diffracted waves or Rayleigh anomaly. The crossing at 1.95 eV (635.9 nm) and 2.33 eV (532.1 nm) are signatures of the SLR and LSPR, respectively. To verify the predictions, we performed a numerical simulation of the scattering spectra for such geometries using Comsol Multiphysics with TE polarization and Floquet boundary conditions. Figure 1c shows the simulated reflectance of the nanodisk lattices as shown in Fig. \ref{fig:fig1}b. The narrowest resonance at 1.95 eV (peak A) only emerges for the lattice spacing $p_2$ = 425 nm, and a kink at $E$ = 2.0 eV is the position of the Rayleigh anomaly. Figures 1d and e confirm the characteristics of the SLR, whose electric field is expanding to the edges of the unit cell. The broad resonant peak at 2.15 eV (peak C) and 2.33 eV (peak B) correspond to the LSPRs of the particles as $p_1$ = 325 nm and $p_2$ = 425 nm, respectively. The electric fields of LSPRs are illustrated in Figs. \ref{fig:fig1}f-g and Figs. \ref{fig:fig1}h-i. The peaks are in good agreement with the CDA predictions shown in Fig. \ref{fig:fig1}b. 

\begin{figure}[htbp]
\centering
\includegraphics[width=0.95\linewidth]{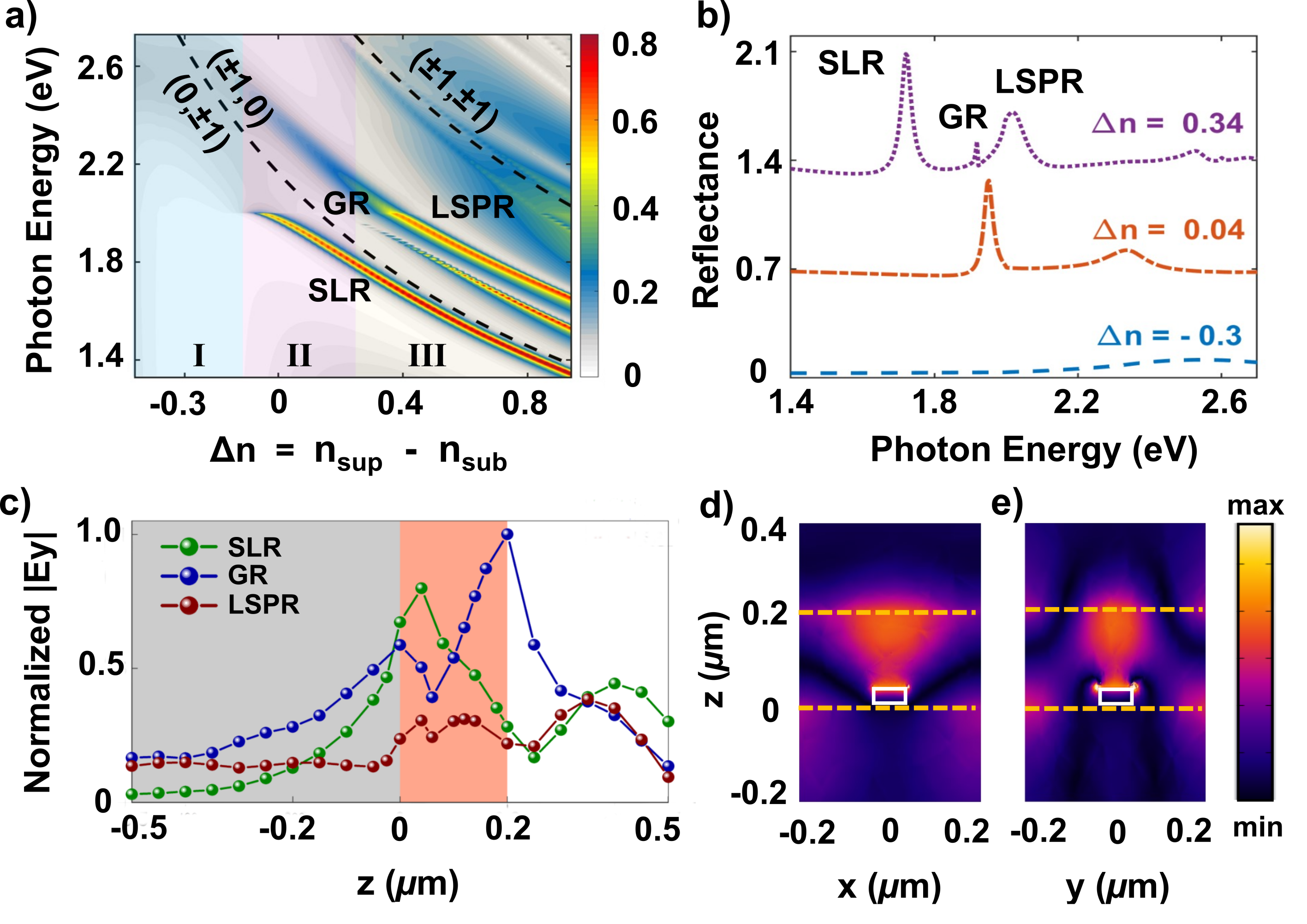}
\caption{Excitations of the SLR with varying refractive index contrast. a) Evolution of the resonances in the index contrast $\Delta n = -0.46 - 0.94$. Three regions (I, II, and III) are observed. The numbers in parenthesizes indicate the diffracted orders of the lattice. b) Truncated spectra at specific values in Fig. 2a. c) Calculated electric field in y-direction for three resonances as a function of z position for $\Delta n = 0.34$. d), e) Electric field of GR.}
\label{fig:fig2}
\end{figure}

We investigated the role of the superstrate for the nanodisk array with a period $p$ = 425 nm in sustaining both LSPRs and SLRs. $n_{sup}$ is varied from 1.0 (air) to 2.4 (TiO$_2$ \cite{Sun2017} or perovskite film \cite{Lu2020, Dang2020}). This corresponds to a variation in the index contrast, defined by $\Delta n = n_{sup} - n_{sub}$, from -0.46 to 0.94. Although SLRs have been studied in inhomogeneous media \cite{Auguie2010}, only low index contrast (<9$\%$) was considered. Here, the contrast range was extended up to 64$\%$ ($\Delta n = 0.94$). Figure \ref{fig:fig2}a shows three regions with distinct features: the region I, $\Delta n < -0.1$, which only supports the LSPR because the waves propagate in the substrate owing to its higher refractive index. Region II supports both LSPR and SLR as the index contrast is kept at a minimum ($-0.1 < \Delta n < 0.3$). Notably, an additional sharp peak appears in the region III ($\Delta n > 0.3$)  between the LSPR and SLR. This sharp resonance is attributed to a guided resonance (GR) \cite{Zentgraf2009,Monticone2017,Bian18} which is a dielectric guided mode confined within the superstrate and is diffracted into the radiative continuum owing to the lattice effect. Naturally, this GR is less exposed to the nonradiative losses of the metallic particles than the LSPR and SLR, thus, it would exhibit much higher Q-factors. To verify the nature of the resonances, we truncated the spectra at three representative values for each region (see Fig. \ref{fig:fig2}b). The spectrum when $\Delta n = -0.3$ shows a single broad LSPR at 2.54 eV while the one of $\Delta n = 0.04$ exhibits two distinct peaks, (LSPR and SLR) as shown in Fig. \ref{fig:fig1}c. Lastly, the spectrum for $\Delta n = 0.34$ reveals three peaks: SLR (1.72 eV), GR (1.92 eV), and LSPR (2.02 eV). Figure \ref{fig:fig2}c presents the calculated electric field in y-direction for the resonances. As expected, the SLR field is maximum in the lattice plane, whereas the GR field is maximum at the boundary between the superstrate and the air. However, the GR field also exhibits a local maximum near the plasmonic particle. This suggests that this GR is resulted from the hybridization between pure guided mode and LSPR. Figures \ref{fig:fig2}d and e confirm this conclusion, which shows that the electric field is mainly confined within in the superstrate and on the particle. Hybridization is also reflected in th Q-factors of the resonances. When $\Delta n$ increases from 0.04 $\rightarrow$ 0.34, $Q_{SLR}$ decreases from  60 $\rightarrow$ 52, while $Q_{LSPR}$ increases from 17 $\rightarrow$ 28, with $Q_{GR}$ = 232 at $\Delta n = 0.34$. The increase in $Q_{LSPR}$ is attributed to its hybrid with GR, in which a loss exchange occurs between two resonances.  

\begin{figure}[t]
\centering
\includegraphics[width=0.95\linewidth]{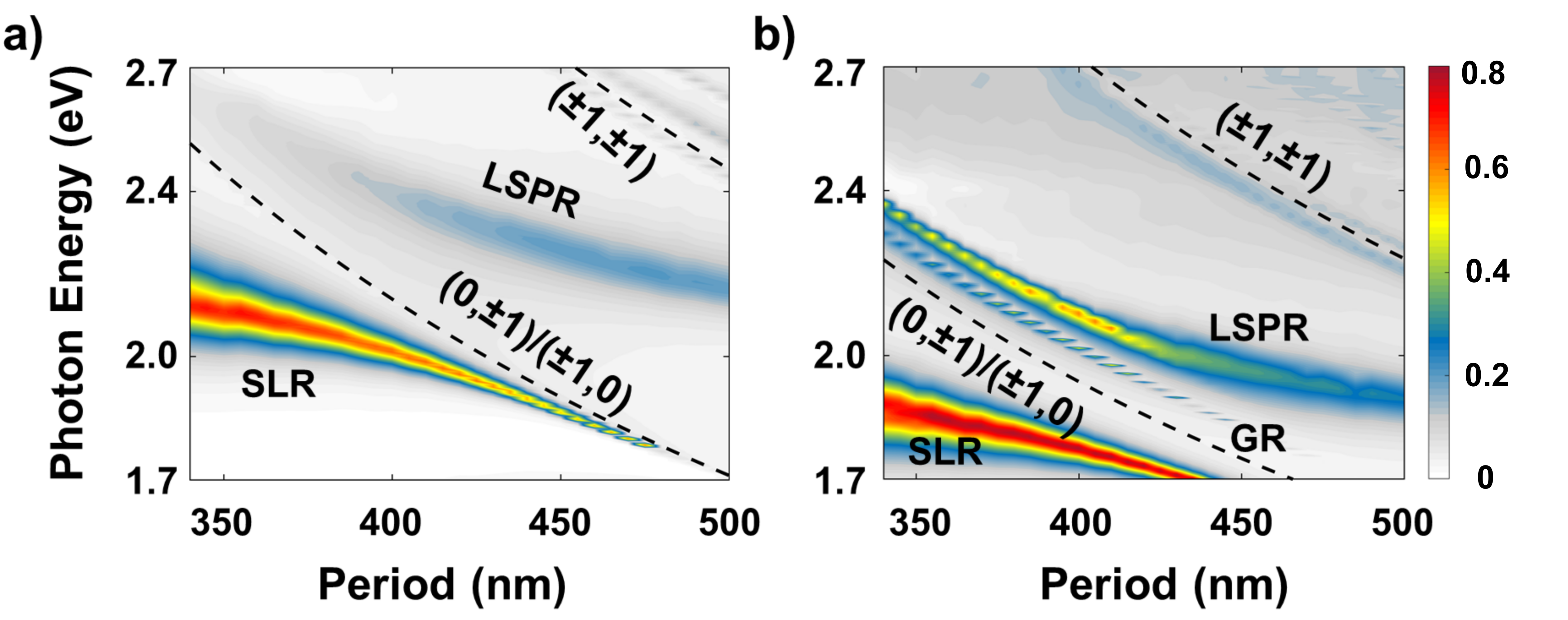}
\caption{The reflectance of the system (at $\theta = 0^{\circ}$) with tuned periodicity for two refractive index contrasts. a) $\Delta n = 0.04$ and b) $\Delta n = 0.34$. The dash curves with numbers in parenthesizes are diffraction orders of the lattice.}
\label{fig:fig3}
\end{figure}

\begin{figure}[htbp]
\centering
\includegraphics[width=0.9\linewidth]{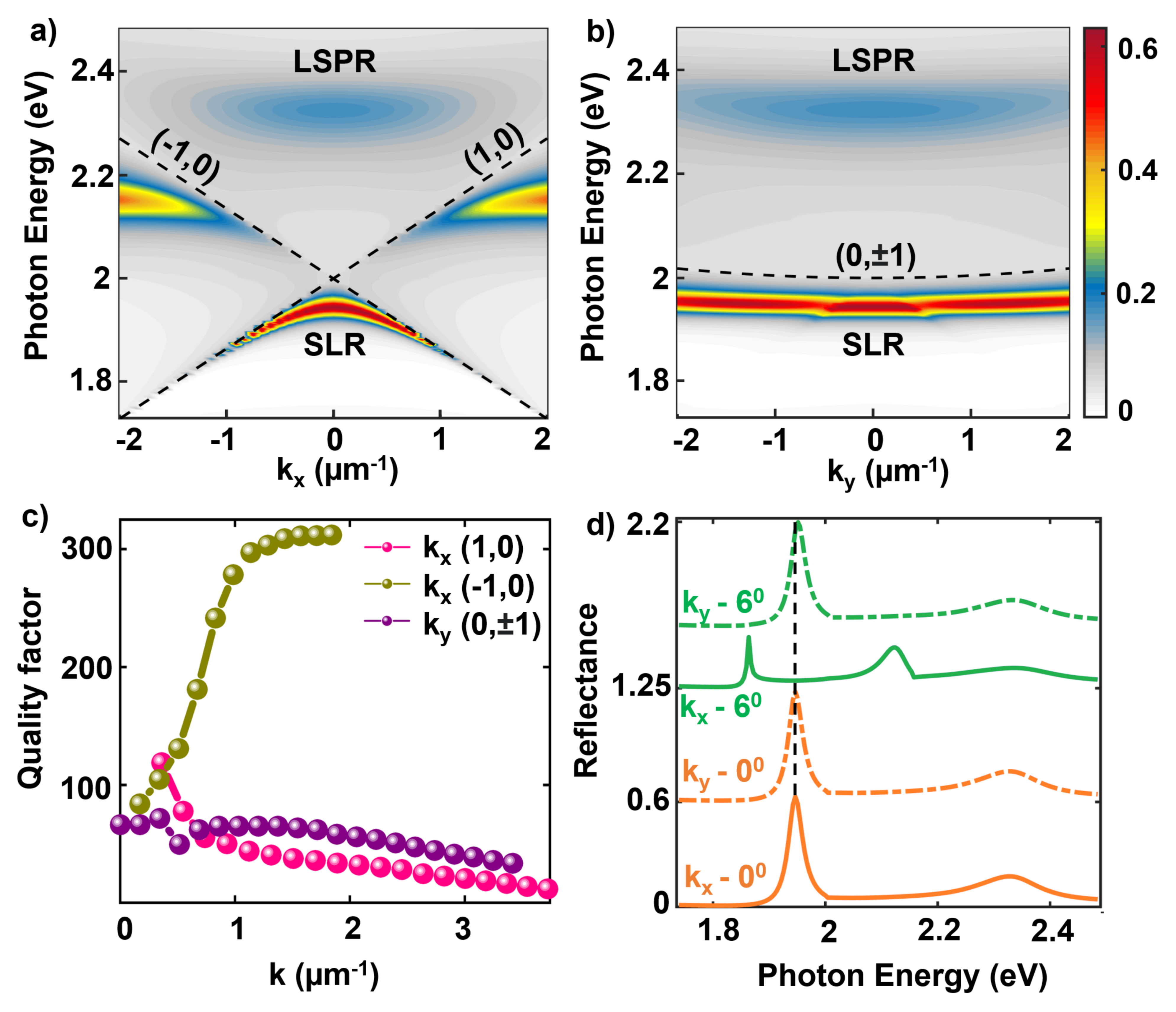}
\caption{Angle-resolved reflectance of the silver nanodisk array for $\Delta n$ = 0.04 and $p$ = 425 nm. The reflectance is probed a) in $k_x$ and b) in $k_y$. c) Q-factor of different bands of SLRs. d) Spectra truncated in a) and b) at $\theta = 0^{\circ}$ and $\theta = 6^{\circ}$. The solid and dashed lines are associated with the spectra probed in $k_x$ and $k_y$, respectively.}
\label{fig:fig4}
\end{figure}

We investigated the influence of the spatial distance between the particles on the optical responses of the system. We focused on two representative cases: low ($\Delta n = 0.04$) and high ($\Delta n = 0.34$) refractive index contrasts, which are presented in Figs. \ref{fig:fig3}a and \ref{fig:fig3}b, respectively. As shown in Fig. \ref{fig:fig3}a, the LSPR frequency decreases with increasing periodicity. This contradicts the common belief that the frequency is solely determined by the materials and the optical environment. The change in the frequency of the LSPR results from the reshaping of the band diagram as the periodicity is tuned. We also observe that the SLR becomes sharper as the periodicity approaches the diffraction line. This can be attributed to the enhanced coupling between the LSPR and Rayleigh anomalies. Notably, under such a large index contrast $\Delta n = 0.34$, the GR is not always excited. Instead, the GR is excited when it couples with the LSPRs. As a result, the losses of the LSPR are alleviated via hybridization with the GR, and hence, the Q-factors of these LSPRs significantly increase. Conversely, the Q-factors of the SLRs deteriorate owing to reduced coupling with the in-plane diffractions. 
We analyzed the angle dependence of the SLRs when excited in a quasi-homogeneous environment, $\Delta n = 0.04$. As they are probed in wavevector $k_{x}$ (Fig. \ref{fig:fig4}a), their optical responses are shaped by the coupling between the LSPR and the diffraction orders  ($\pm$1,0). The LSPR in Figs. \ref{fig:fig4}a and b exhibits relatively constant energy when excited from oblique angles. In contrast, the SLRs emerging from the coupling between the LSPR and the diffraction order ($\pm$1,0) show strong angular dependence. Two SLR branches can be observed along $k_x$, both  exhibiting a linear dispersion as they approach the diffraction order ($\pm$1,0). The upper SLR branch fades closer to the $\Gamma$ point and is not visible at normal incidence. In comparison, the lower SLR branch has a pronounced resonance signal at the $\Gamma$ point and fades at oblique angles. When probed in $k_y$, only the lower SLR branch is visible with a quadratic dispersion, closely following the characteristics of the diffraction order (0,$\pm$1). We note that changing polarization and fixing the probe angle will produce the same results as a fixed polarization and changing probing directions as in this work. Figure \ref{fig:fig4}c presents the Q-factors of the different SLR branches plotted as a function of the wavevector. As $k_x$ deviates from the $\Gamma$ point, the Q-factor increases rapidly for the SLR branch (-1,0) and reaches a plateau (Q = 300) when fading to the Rayleigh anomalies ($k_x > 1 \mu m^{-1}$).

\begin{figure}[htbp]
\includegraphics[width=0.95\linewidth]{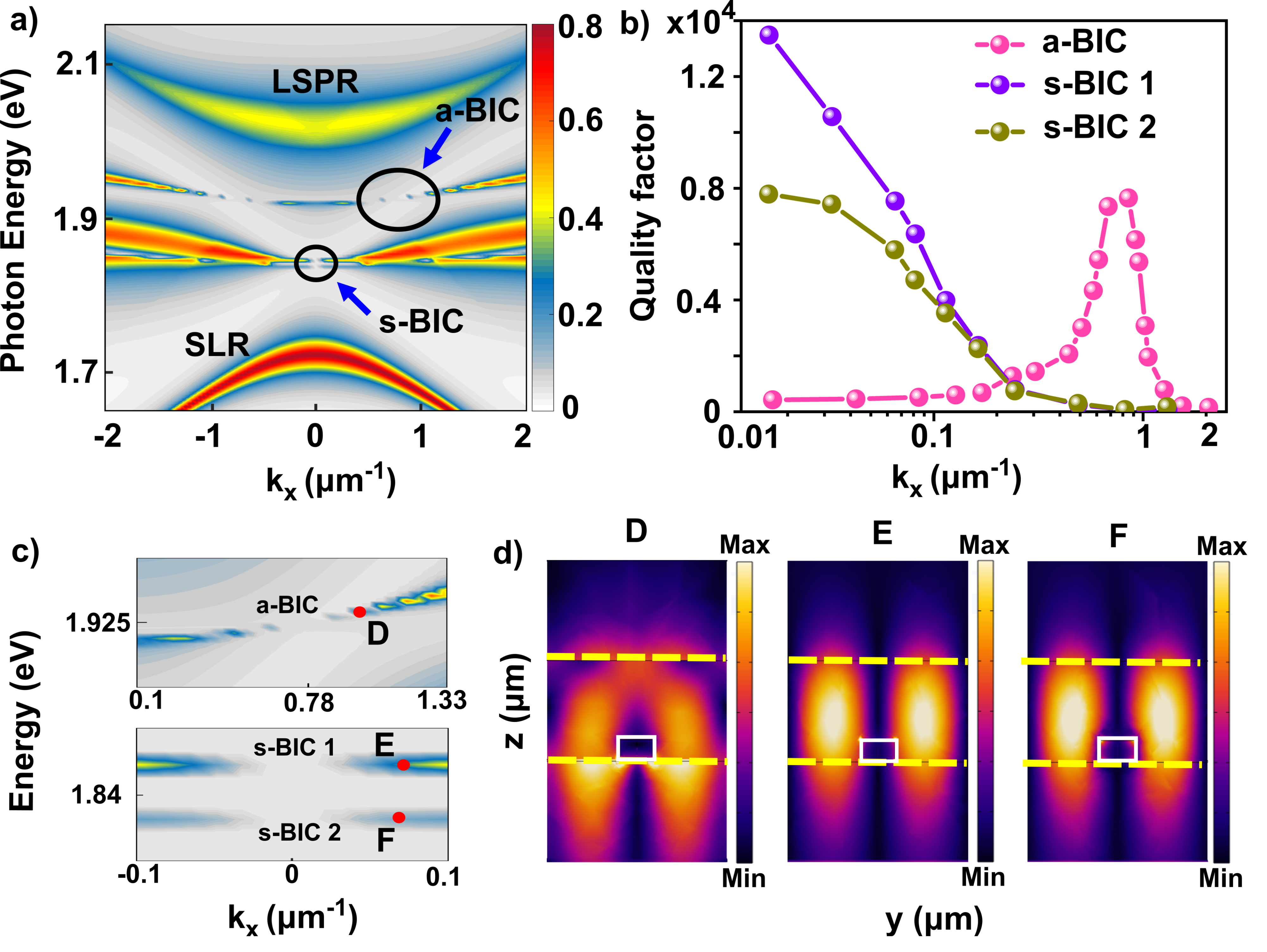}
\caption{Angle-resolved reflectance when $\Delta n$ = 0.34: a) Reflectance spectra are probed along $k_{x}$. Besides LSPR and SLR, the structures form an accident BIC (a-BIC) at $k_{x} = 0.78~\mu m^{-1}$ and two quasi symmetry-protected BICs (s-BICs) at $\Gamma$ point. b) Total Q-factors of the BICs marked in a). c) Magnified reflectance around quasi-BIC points indicated in a). The color scale is unchanged compared to that in a). d) Side view of the electric fields of points D, E, and F in c). They confirm that the a-BIC is an hybrid LSPR-GR state while the two s-BICs are both only a GR nature.}
\label{fig:fig5}
\end{figure}

Finally, we investigated the angle-resolved response of the structure with a high-index contrast $\Delta n$ = 0.34. As shown in Fig. \ref{fig:fig5}, we find strikingly different responses compared with the low-index contrast case, as shown in Fig. \ref{fig:fig4}. First, the LSPR and lower SLR bands are reshaped by the contraction of the band diagram owing to the change in the refractive index. This leads to a quadratic dispersion of the LSPR bands. More importantly, three GR-like bands with a high curvature emerged between the LSPR and lower SLR bands. The highest energy band corresponds to the one denoted as $GR$ in our previous studies (Figs. \ref{fig:fig2} and \ref{fig:fig3}), resulting from the hybridization between a pure GR and the LSPR. Here, the angular-resolved reflectance of this band reveals that it gets sharper when moving away from the G point and eventually reaches its highest quality factor at $k_{x} = 0.78\,\mu m^{-1}$ and $E$ = 1.925 eV ($\Lambda$ = 675.3 nm) when its scattering resonance vanishes locally in the momentum space, see Fig. \ref{fig:fig5}b. Such behaviors are the hallmarks of an accidental BIC (a-BIC) that occurs when the hybridization mechanism leads to a destructive interference configuration \cite{Hsu2016}.
On the other hand, the spectral range and curvature of the second and third GR-like bands suggest that they are hybridizations between a pure GR and the upper SLR branch depicted in Fig. \ref{fig:fig5}a. Moreover, because the upper SLR branch does not exist in the vicinity of the $\Gamma$ points (i.e. $|k_x|<0.1\,\mu m^{-1}$, Fig. \ref{fig:fig5}a), these two bands are of a GR nature solely in this range. Interestingly, their scattering resonances vanish at the $\Gamma$ point when energies $E$ = 1.845 eV ($\Lambda$ = 672 nm) and 1.836 eV ($\Lambda$ = 675.3 nm), while becoming broader as soon as they are under oblique excitation. Therefore, they are symmetry-protected BICs (s-BICs), in which the coupling to the radiative continuum is strictly forbidden owing to the $C_4$ symmetry mismatch between their in-plane pattern and the radiating plane waves \cite{Hsu2016}.Magnified views of the a-BIC and the two s-BICs are presented in the upper and lower panels of Fig. \ref{fig:fig5}c. Figure \ref{fig:fig5}d depicts the electric field distributions of the three BIC states, confirming that the a-BIC is indeed of a hybrid GR–LSPR nature, while the two s-BICs are solely of a GR nature. We used the Q-factor to compare these BICs with others reported for plasmonic systems. The Q-factors in our system reached remarkably high values ($\sim10^4$) in the BIC states, as shown in Fig. \ref{fig:fig5}b. We note that in plasmonic systems, the Q-factor is bounded by the nonradiative losses from the metallic component, and most of the reported BICs from plasmonic systems designed for infared and visible frequencies exhibit Q-factors only in the range of $10^2-10^3$ \cite{Azzam2018,Liang2020,Seo}. Our results show that by using GR-like modes, the lossy flaw in plasmonic systems can be mitigated, thus providing ultra-high-quality factors even with metallic components. A similar result was recently reported for 1D plasmonic gratings  \cite{Deng2021}.



In conclusion, we demonstrated that plasmonic structures can enable a variation in the resonance as well as in the trapped light in the form of BICs. The key constraint to unlock this interesting phenomenon is the large refractive index contrast surrounding the plasmonic particles. Our study demonstrated a new application of guided resonance as an effective mechanism to mitigate the absorptive losses of the plasmonic particles and revealed that the waveguide effect can lead to the formation of BICs in a lossy medium. The findings could be implemented in plasmonic lattice covered by a thin film of perovskite or semiconductor nanocrystals for applications in sensing and light–emitting devices.\\

\noindent\textbf{Funding Information}
This work was funded by Vingroup Big Data Institute, Vingroup VINIF.2021.DA00169.\\

\noindent\textbf{Acknowledgement.} Q. L-V would like to thank Bojorn Maes for his critical assistance with simulation models.\\

\noindent\textbf{Disclosures.} The authors declare no conflicts of interest.
\bibliography{Biblio}
\end{document}